%% file: ms.tex
\documentclass{article}
\usepackage{waspaa23,amsmath,graphicx,url,times}
\usepackage{color}
\usepackage{amsmath,amsfonts,amssymb,physics}
\usepackage{algorithm}
\usepackage{array}
\usepackage{textcomp}
\usepackage{stfloats}
\usepackage{float}
\usepackage{verbatim}
\usepackage{setspace}
\usepackage[noend]{algpseudocode}

\usepackage{caption}
\usepackage{subcaption}
\usepackage{xcolor}

\usepackage{comment}
\usepackage{cite}
\usepackage{soul}
\usepackage{epstopdf}
\usepackage{xfrac}
\usepackage{pifont}

\usepackage{cleveref}
\usepackage{csquotes}
\usepackage{booktabs}
\usepackage{multirow}
\usepackage{tabularx,ragged2e}
\newcolumntype{Y}{>{\centering\arraybackslash}X}
\usepackage{footmisc}

\usepackage{varwidth}
\usepackage{tikz}
\usepackage{pgf}
\usepackage{pgfplots}
\usepackage{pgfplotstable}
\pgfplotsset{compat=1.3}
\usetikzlibrary{shapes.geometric}
\usetikzlibrary{shapes.arrows}
\usepackage{array}
\usepackage[nolist, nohyperlinks]{acronym}

\newcommand{\D}[1]{\mathrm{d}{#1}}
\newcommand{\state}[0]{\mathbf{X}_\tau}
\newcommand{\wien}[0]{\mathbf{W}_\tau}
\newcommand{\clean}[0]{\mathbf{X}_0}
\newcommand{\reverb}[0]{\mathbf{Y}}
\newcommand{\sco}[0]{\nabla_{\state} \log p(\state)}
\newcommand{\Post}[0]{\widehat{\mathbf M}(\state)}
\newcommand{\post}[0]{\widehat{\mathbf m}(\state)}

\newcommand{\santiv}{\vspace{-0.5em}}

\definecolor{statedps}{HTML}{FFC300}
\definecolor{dps}{HTML}{EC610A}
\definecolor{kodrasi}{HTML}{000000}

\title{Diffusion Posterior Sampling for Informed Single-Channel
Dereverberation}

\name{Jean-Marie Lemercier$^{1}$\sthanks{Funded by the Federal Ministry for Economic Affairs and Climate Action, project 01MK20012S, AP380. Corresponding author: \texttt{jeanmarie.lemercier@uni-hamburg.de}},
      Simon Welker$^{1,2}$\sthanks{Funded by DASHH (Data Science in Hamburg Hemholtz Graduate School for the Structure of Matter) with the Grant No. HIDSS-0002.},
      Timo Gerkmann$^{1}$}
\address{$^1$ Signal Processing (SP), Universität Hamburg, Germany\\
         $^2$  Center for Free-Electron Laser Science, DESY, Hamburg, Germany\\
}

\begin{document}

\begin{acronym}
\acro{stft}[STFT]{short-time Fourier transform}
\acro{istft}[iSTFT]{inverse short-time Fourier transform}
\acro{dnn}[DNN]{deep neural network}
\acro{pesq}[PESQ]{Perceptual Evaluation of Speech Quality}
\acro{polqa}[POLQA]{perceptual objectve listening quality analysis}
\acro{wpe}[WPE]{weighted prediction error}
\acro{psd}[PSD]{power spectral density}
\acro{rir}[RIR]{room impulse response}
\acro{snr}[SNR]{signal-to-noise ratio}
\acro{lstm}[LSTM]{long short-term memory}
\acro{polqa}[POLQA]{Perceptual Objectve Listening Quality Analysis}
\acro{sdr}[SDR]{signal-to-distortion ratio}
\acro{estoi}[ESTOI]{Extended Short-Term Objective Intelligibility}
\acro{elr}[ELR]{early-to-late reverberation ratio}
\acro{tcn}[TCN]{temporal convolutional network}
\acro{rls}[RLS]{recursive least squares}
\acro{asr}[ASR]{automatic speech recognition}
\acro{ha}[HA]{hearing aid}
\acro{ci}[CI]{cochlear implant}
\acro{mac}[MAC]{multiply-and-accumulate}
\acro{vae}[VAE]{variational auto-encoder}
\acro{gan}[GAN]{generative adversarial network}
\acro{tf}[T-F]{time-frequency}
\acro{sde}[SDE]{stochastic differential equation}
\acro{ode}[ODE]{ordinary differential equation}
\acro{drr}[DRR]{direct to reverberant ratio}
\acro{lsd}[LSD]{log spectral distance}
\acro{sisdr}[SI-SDR]{scale-invariant signal to distortion ratio}
\acro{mos}[MOS]{mean opinion score}
\acro{map}[MAP]{maximum a posteriori}
\acro{rtf}[RTF]{real-time factor}
\end{acronym}

\ninept
\maketitle

\begin{abstract}
We present in this paper an informed single-channel dereverberation method based on conditional generation with diffusion models. With knowledge of the room impulse response, the anechoic utterance is generated via reverse diffusion using a measurement consistency criterion coupled with a neural network that represents the clean speech prior.
The proposed approach is largely more robust to measurement noise compared to a state-of-the-art informed single-channel dereverberation method, especially for non-stationary noise.
Furthermore, we compare to other blind dereverberation methods using diffusion models and show superiority of the proposed approach for large reverberation times.
We motivate our algorithm by introducing an extension for blind dereverberation allowing joint estimation of the room impulse response and anechoic speech.
Audio samples and code can be found online\footnote{https://uhh.de/inf-sp-derev-dps}.
\end{abstract}


\santiv
\section{Introduction}
\santiv
\input{sections/intro}

\santiv
\section{Diffusion-based generative models}
\santiv
\input{sections/sgm}

\santiv
\santiv
\section{Diffusion Posterior Sampling for Dereverberation}
\santiv
\input{sections/dps}

\santiv
\section{Experimental Setup}
\santiv
\input{sections/exp}

\santiv
\section{Results and Discussion}
\santiv
\input{sections/results}

\santiv
\section{Conclusions and Future Work}
\santiv
\input{sections/conclusion}

\bibliographystyle{IEEEtran}
\santiv
\bibliography{refs23}

\end{document}

%% file: sections/intro.tex
Reverberation is a natural phenomenon occurring in most spaces of our daily life, where sound waves get reflected and attenuated by the enclosure walls. It degrades speech intelligibility and quality for normal listeners, and dramatically so for hearing-impaired listeners \cite{Naylor2011}. Therefore, modern communication devices and listening setups are equipped with dereverberation algorithms which aim to recover the anechoic component of speech~\cite{Naylor2011}. We will denote as \textit{informed} the methods that exploit prior knowledge of the \ac{rir} and as \textit{blind} the methods that try to recover anechoic speech without knowing the \ac{rir}.

Traditional blind dereverberation methods exploit the statistical properties of the anechoic and reverberant signals, typically in the time, spectral or cepstral domain~\cite{gerkmann2018book_chapter}. Machine learning techniques try to learn these statistical properties directly from data \cite{wang2018supervised}. Typically, supervised predictive models for blind dereverberation include \ac{tf} maskers~\cite{Williamson2017j}, time domain methods~\cite{Ernst2019} and direct spectro-temporal mapping \cite{Han2017}. Generative models, that aim to learn the posterior distribution of clean speech conditioned on corrupted speech, have also been introduced for blind dereverberation and speech enhancement. In particular, conditional diffusion-based generative models (or simply \textit{diffusion models}) \cite{ho2020denoising, song2021sde}
have been successfully applied to blind dereverberation \cite{Richter2022SGMSE++, Lemercier2022storm, lu2022conditional}.

Though informed dereverberation may seem an easier task in comparison to blind dereverberation, knowing the \ac{rir} does not guarantee to find a stable and causal inverse filter in the single-channel case, as typical real-world RIRs are mixed-phase signals \cite{Neely1979Invertibility}. Using multiple microphones may mend such issues to some extent \cite{Miyoshi1988MINT}, but may also suffer from limited robustness \cite{Hikichi2007Inverse}.
Single-channel informed methods include least-squares and $\mathcal{L}^p$-based optimization rules \cite{Mourjopoulos1982Comparative, Mertins2010Shortening, Schepker2022Robust}, 
frequency-domain methods such as homomorphic inverse filtering \cite{Mourjopoulos1982Comparative}
, and hybrid techniques such as \cite{Kodrasi2014} where a regularized inverse filter is used to avoid non-causality artifacts and a speech enhancement scheme is used as a post-processing step to attenuate residual pre-echoes.

In this paper, we present a single-channel informed dereverberation technique using diffusion models, with two variants for reverse sampling. We show that the proposed method retrieves high-quality anechoic speech samples for all reverberant conditions without the need for post-processing. We also demonstrate the robustness of the proposed method to measurement noise. We compare our results with a state-of-the-art frequency-domain informed dereverberation method \cite{Kodrasi2014} as well as recently introduced diffusion models for blind dereverberation \cite{Richter2022SGMSE++, Lemercier2022storm}. 
Code and audio examples are provided in the supplementary material.

%% file: sections/sgm.tex
In this section we introduce diffusion models, a class of generative models that has recently showed impressive abilities to learn natural data distributions in the image \cite{ho2020denoising, song2021sde} and speech domains \cite{lu2022conditional, Welker2022SGMSE, Richter2022SGMSE++}.
Score-based diffusion models in the framework by Song et al. \cite{song2021sde} are defined by three  components: a forward diffusion process parameterized by a \ac{sde}, a score estimator implemented by a \ac{dnn} and a sampling method for inference.

As in \cite{Welker2022SGMSE,Richter2022SGMSE++,Lemercier2022analysing,Lemercier2022storm}, here the processes are defined in the complex spectrogram domain, independently for each \ac{tf} bin. In the following, the variables in uppercase bold are assumed to be vectors $\mathbb C^D$ containing coefficients of a flattened complex spectrogram--- with $D$ the product of the time and frequency dimensions--- whereas variables in lowercase bold are time vectors in $\mathbb{R}^L$ (unless specified) and variables in regular font are scalars in $\mathbb{C}$.
The stochastic \textit{forward process} $\{\state\}_{\tau=0}^T$ slowly transforms clean speech into a tractable noise distribution
. It is modeled as the solution to the following Variance-Exploding \ac{sde} \cite{song2021sde
}: 
\begin{equation} \label{eq:forward-sde}
    \D{\state} = g(\tau) \D{\wien},
\end{equation}
\begin{equation}
g(\tau) = \sigma_\mathrm{min} \left( \frac{\sigma_\mathrm{max}}{\sigma_\mathrm{min}}\right)^\tau \sqrt{2 \log \frac{\sigma_\mathrm{max}}{\sigma_\mathrm{min}}},
\end{equation}
where $\state$ is the current state of the process indexed by a continuous time variable $\tau \in [0, T]$. The stochastic process $\wien$ is a standard $D$-dimensional Brownian motion, which implies that $\D{\wien}$ is a zero-mean Gaussian random variable with infinitesimal standard deviation for each \ac{tf} bin. 
The initial condition $\clean = \mathbf X$ represents clean speech and the \textit{diffusion coefficient} $g$ controls the amount of white noise injected at each step, with $\sigma_\mathrm{min}$ and $\sigma_\mathrm{max}$ being hyperparameters representing extremal noise levels.

The \textit{reverse process} $\{\state\}_{\tau=T}^0$ turning noise into clean speech is another diffusion process also defined as the solution of a \ac{sde} \cite{anderson1982reverse, song2021sde}, with $\tau$ flowing in reverse (i.e. $\D{\tau}<0$). Here, we will use the corresponding \textit{probability flow} \ac{ode}, since its solution has the same marginal distribution as its \ac{sde} counterpart \cite{song2021sde}:
\begin{equation}\label{eq:reverse-ode}
    \D{\state} = 
            - \frac{1}{2} g(\tau)^2 \sco \D{\tau}.
\end{equation}

The quantity $\sco$ is the \textit{score function}, i.e. the gradient of the logarithm distribution for the current state $\state$.
At inference time, this score function is not available, and therefore a neural network $s_\theta(\state, \sigma(\tau))$, called \textit{score model}, is used to estimate the score of the current state $\state$ given the current Gaussian noise standard deviation $\sigma(\tau)$. 
The latter
encodes 
how much Gaussian noise is left to remove before getting in the vicinity of clean speech $\clean$. It must therefore be fed to the score network as conditioning, and is obtained in closed-form for the Variance-Exploding SDE\cite{song2021sde}.
The score model $s_\theta$ is trained via \textit{denoising score matching}~\cite{vincent2011connection}.

%% file: sections/dps.tex
\subsection{Diffusion Posterior Sampling for Inverse Problems}

Inverse problems consist in finding the state $\mathbf X$ given a observation $\reverb = \mathcal{A}(\mathbf X)$ with $\mathcal{A}$ being a measurement operator. We consider the \textit{non-blind noisy linear} inverse problem of informed single-channel dereverberation. That is, we wish to retrieve the anechoic version of some reverberant speech under measurement noise $\mathbf n$ when the \ac{rir} $\mathbf k \in \mathbb{R}^K$ is known. We define the mixing process in the time-domain, with $\mathbf x := \mathrm{iSTFT}(\mathbf X)$ and $\mathbf y := \mathrm{iSTFT}(\reverb)$, as: 
\begin{equation} \label{eq:measurement}
    \mathbf y = \mathbf k \ast \mathbf x + \mathbf n \text{ with } \mathbf n \sim \mathcal{N}(0, \eta^2),
\end{equation}
where iSTFT denotes inverse short-time Fourier transformation and $\ast$ is the time-domain linear convolution resulting in $\mathbf y \in \mathbb{R}^{L+K-1}$.

Diffusion posterior sampling (DPS) is a technique based on diffusion models that was proposed for solving inverse problems \cite{chung_diffusion_2022} and was recently applied to music restoration tasks \cite{moliner_solving_2022}. 
The score function is used as a surrogate speech prior and a log-likelihood term is added to the reverse diffusion, so that the output sample belongs to the posterior $p(\mathbf X|\reverb)$. The unconditional score $\sco$ in \eqref{eq:reverse-ode} is then replaced by the score of the posterior, in order to include the measurement model in the sampling process:
\begin{equation} \label{eq:posterior-decompo}
\nabla_{\state} \log p(\state|\reverb) = \sco + \nabla_{\state} \log p(\reverb|\state).
\end{equation}
For sampling, a trained score model $s_\theta(\state, \sigma(\tau)) \approx \sco$ is needed, as well as an approximation of the log-likelihood gradient $\nabla_{\state} \log p(\reverb|\state)$, since it is generally intractable.

\input{plots/algorithm}
\santiv
\subsection{Log-likelihood approximation}
\santiv

\noindent \textit{a) Posterior mean approximation}: \vspace{0.5em}

In \cite{chung_diffusion_2022}, the log-likelihood approximation is carried by transferring the outer marginalization with regard to $\clean$ inside the conditioning, thereby assuming that the \textit{posterior mean} $\mathbb{E}[\clean|\state]$ is a sufficient statistic for $\state$ when modelling the likelihood function:
\begin{align}
    p(\reverb|\state) 
    &= \int p(\reverb|\clean) p(\clean|\state) \D{\clean} \nonumber \\
    &\approx p(\reverb| \underbrace{\int \clean p(\clean|\state) \D{\clean}}_{\mathbb{E}[\clean|\state]} ).
\end{align}

The posterior mean is obtained via the Tweedie formula \cite{Efron2011}, and can be approximated using our score function estimator :
\begin{align}
    \Post 
    &=  \state + \sigma^2(\tau) \sco \label{eq:ideal_denoiser} \\
    & \approx \state + \sigma^2(\tau) s_\theta(\state, \sigma(\tau)).
\end{align}

Our measurement model \eqref{eq:measurement} yields the following \textit{posterior mean approximation} for the log-likelihood gradient:
\begin{equation}\label{eq:dps-likelihood}
\nabla_{\state} \log p(\reverb |\state) \approx - \frac{1}{\eta^2} \nabla_{\state} || \mathbf y - \mathbf k \ast \post||_2^2,
 \end{equation}
 where $\post := \mathrm{iSTFT}(\Post)$ and $\eta$ is the measurement noise level.
The resulting reverse probability flow ODE is:
\begin{equation} \label{eq:dps-reverse}
   \hspace{-0.2em} \resizebox{0.9\hsize}{!}{%
        $\D{\state} = - \frac{1}{2} g(\tau)^2 s_\theta(\state, \sigma(\tau)) \D{\tau} + \zeta(\tau, \eta) \nabla_{\state} || \mathbf y - \mathbf k \ast \post||_2^2 \D{\tau}$,}
\end{equation}
with $\zeta(\tau, \eta) > 0$ a hyperparameter controlling the importance of the measurement error term. According to \eqref{eq:posterior-decompo}, its theoretical value should be $\zeta(\tau, \eta) = g(\tau)^2 / ( 2 \eta^2 )$. In \cite{chung_diffusion_2022} however, this hyper-parameter is empirically set to $\zeta(\tau, \eta) = \zeta^{'}(\tau) / || \mathbf y - \mathbf k \ast \post||_2$ so that the measurement error magnitude itself does not influence the importance of the gradient step, and $\zeta^{'}(\tau)$ is a schedule which we will describe later in Section \ref{sec:exp:hyperparameters}.
\vspace{0.5em}

\noindent \textit{b) State approximation}:  \vspace{0.25em}

In \cite{shoushtari2022dolph
}, a different approximation is used, where the measurement model \eqref{eq:measurement} takes as clean speech reference the current state $\state$ itself, rather than the posterior mean $\Post$. This yields the following \textit{state approximation} for the log-likelihood gradient:
\vspace{-0.5em}
\begin{equation} \label{eq:dolph-likelihood}
     \nabla_{\state} \log p(\reverb|\state) \approx  - \frac{1}{\eta^2} \nabla_{\state} || \mathbf y - \mathbf k \ast \mathbf x_\tau||_2^2,
\end{equation}
with $\mathbf x_\tau := \mathrm{iSTFT}(\state)$. In turn, this results in the following reverse probability flow ODE:
\begin{equation} \label{eq:dolph-reverse}
   \hspace{-0.1em} \resizebox{0.9\hsize}{!}{%
        $\D{\state} = - \frac{1}{2} g(\tau)^2 s_\theta(\state, \sigma(\tau)) \D{\tau} + \zeta(\tau, \eta) \nabla_{\state} || \mathbf y - \mathbf k \ast \mathbf x_\tau||_2^2 \D{\tau}$,}
\end{equation}
\noindent this time with $\zeta(\tau, \eta) = \zeta^{'}(\tau) / || \mathbf y - \mathbf k \ast \mathbf{x}_\tau||_2$. This approximation becomes less valid as the noise level $\sigma(\tau)$ increases, since for large noise levels, the state $\mathbf{x}_\tau$ is a much worse estimate of $\mathbf{x}_0$ compared to the posterior mean $\post$.
In practice, the reverse probability flow ODEs \eqref{eq:dps-reverse} and \eqref{eq:dolph-reverse} are solved using a predictor-corrector numerical scheme \cite{song2021sde} (see Section \ref{sec:exp:hyperparameters}).

\input{plots/graph_t60}

\input{plots/lineplot_measerr}

%% file: plots/algorithm.tex
\newcounter{phase}[algorithm]
\newlength{\phaserulewidth}
\newcommand{\setphaserulewidth}{\setlength{\phaserulewidth}}
\newcommand{\phase}[1]{%
  \vspace{0.1em}
  \Statex\strut\refstepcounter{phase}
  \hspace{1em}\textit{Phase~\thephase~--~#1}%
  \vspace{-2ex}
  \Statex\leavevmode\llap{ 
  }
  \hspace{1em} 
  \rule{0.9\linewidth}{\phaserulewidth}
}
\makeatother
\setphaserulewidth{.3pt}

\newcommand{\antiv}[0]{\vspace{-0.75em}}
\newcommand{\bigantiv}[0]{\vspace{-1.75em}}

\begin{algorithm}
\caption{Posterior Sampling Scheme
}\label{alg:inference}
\hspace*{\algorithmicindent} \textbf{Input}: Corrupted $\reverb$, \ac{rir} $\mathbf k$, 
Reverse step size $\Delta \tau = - \frac{T}{N}$ \\
 \hspace*{\algorithmicindent} \textbf{Output}: Clean speech estimate $\widehat{X}$
\begin{algorithmic}[1]
\setstretch{1.}
\State Sample initial reverse state $\mathbf X_T  \sim \mathcal{N}(0, \mathbf{I})$

\For{$n \in \{ N, \dots, 1 \} $}

\State Get diffusion time $\tau = n \frac{T}{N}$

\phase{Corrector}
\State Estimate score $s_\theta(\state, \sigma(\tau))$
\State Sample correction noise $\mathbf W_c \sim \mathcal{N}(0, \mathbf{I})$
\State Correct estimate: 
\antiv
\begin{equation*}
    \state \leftarrow \state + 2 r^2 \sigma(\tau)^2 s_\theta(\state, \sigma(\tau)) + 2 r \sigma(\tau) \mathbf{W}_c 
\end{equation*}
\bigantiv

\phase{Predictor}
\State Estimate score $s_\theta(\state, \sigma(\tau))$
\State Predict next Euler step: 
\antiv
\begin{equation*}
\state \leftarrow \state - \frac{1}{2} g(\tau)^2 s_\theta(\state, \sigma(\tau)) \Delta \tau 
\end{equation*}
\bigantiv

\phase{Posterior}

\State \textcolor{statedps}{\textbf{if} StateDPS \textbf{then} use state approximation \eqref{eq:dolph-likelihood}:} 
\antiv
\begin{equation*}
\textcolor{statedps}{\mathbf x^{(\mathrm{int})} = \mathrm{iSTFT}(\state)}
\end{equation*}
\bigantiv 

\State \textcolor{dps}{\textbf{if} DPS \textbf{then} use posterior mean approximation \eqref{eq:dps-likelihood}:} 
\antiv
\begin{equation*}
\textcolor{dps}{\mathbf x^{(\mathrm{int})} = \mathrm{iSTFT} \left( \Post \right)}
\end{equation*}
\bigantiv

\State Add log-likelihood gradient:
\antiv
\begin{equation*}
\state \leftarrow \state + \zeta(\tau, \eta) \nabla_{\state} || \mathbf y - \mathbf k \ast \mathbf x^{(\mathrm{int})} ||_2^2 \Delta\tau
\end{equation*}
\antiv
\antiv

\EndFor

\State Output estimate: $\widehat{\mathbf{X}} = \clean$

\end{algorithmic}
\end{algorithm}

%% file: plots/graph_t60.tex
\newcommand{\gw}[0]{0.325\textwidth}

\begin{figure*}[t]
    \hspace{3cm}
    \includegraphics[width=0.7\textwidth,trim={0cm 0cm 2.7cm 0.3cm},clip]{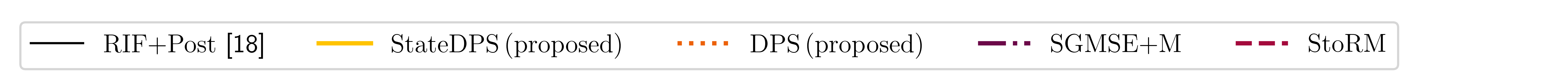}

    \centering
    \includegraphics[width=\gw]{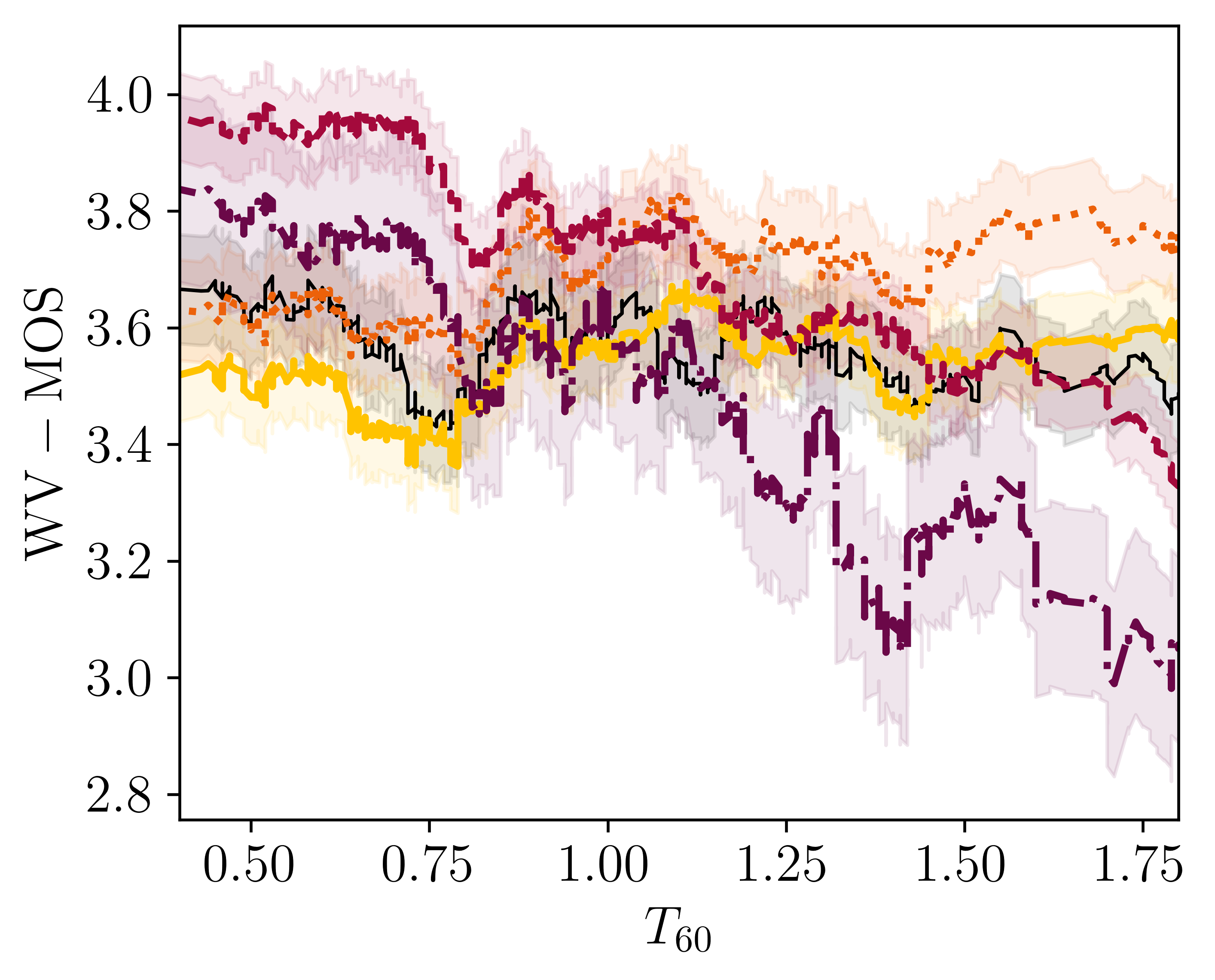}
    \includegraphics[width=\gw]{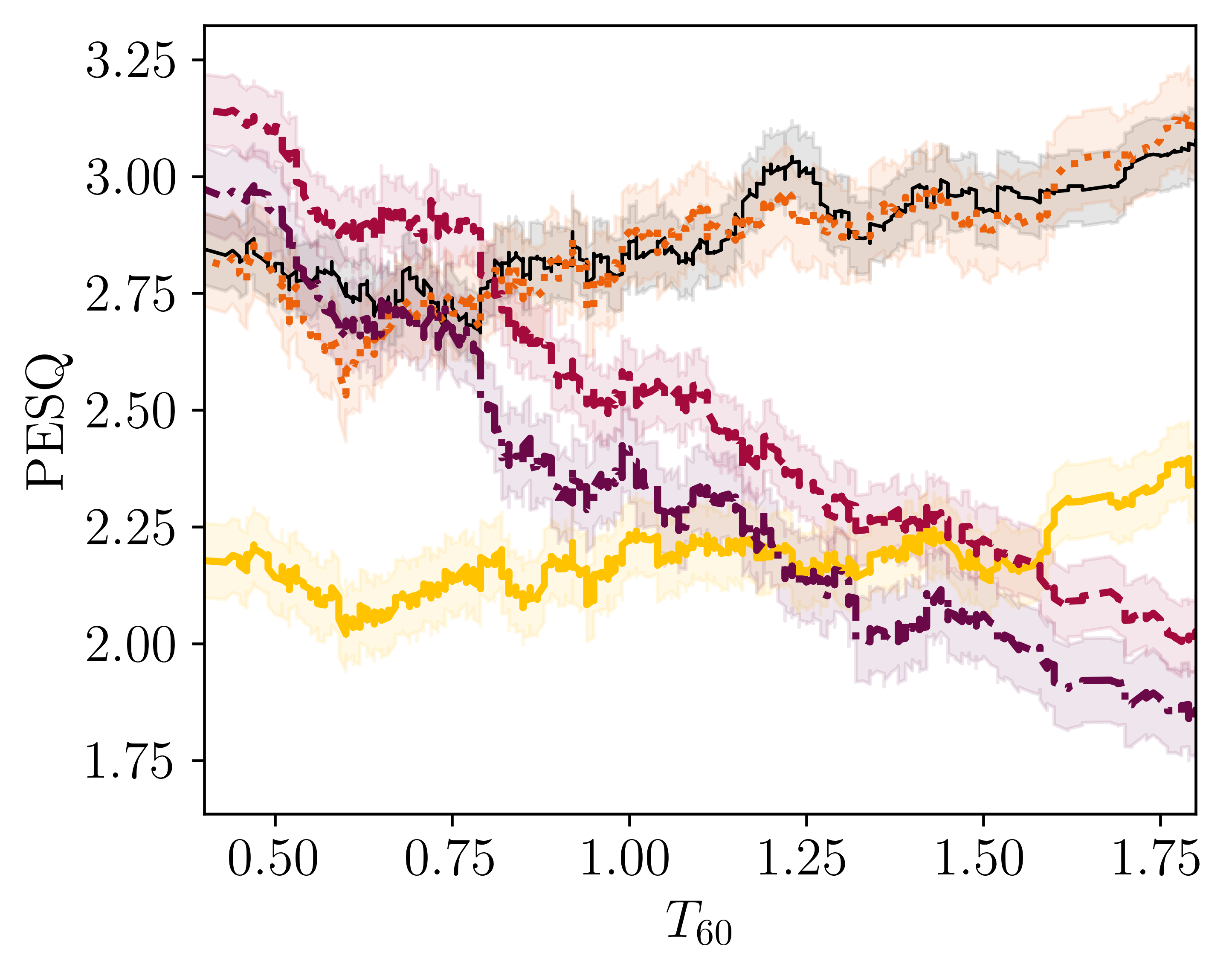}
    \vspace{-1em}
    \includegraphics[width=\gw]{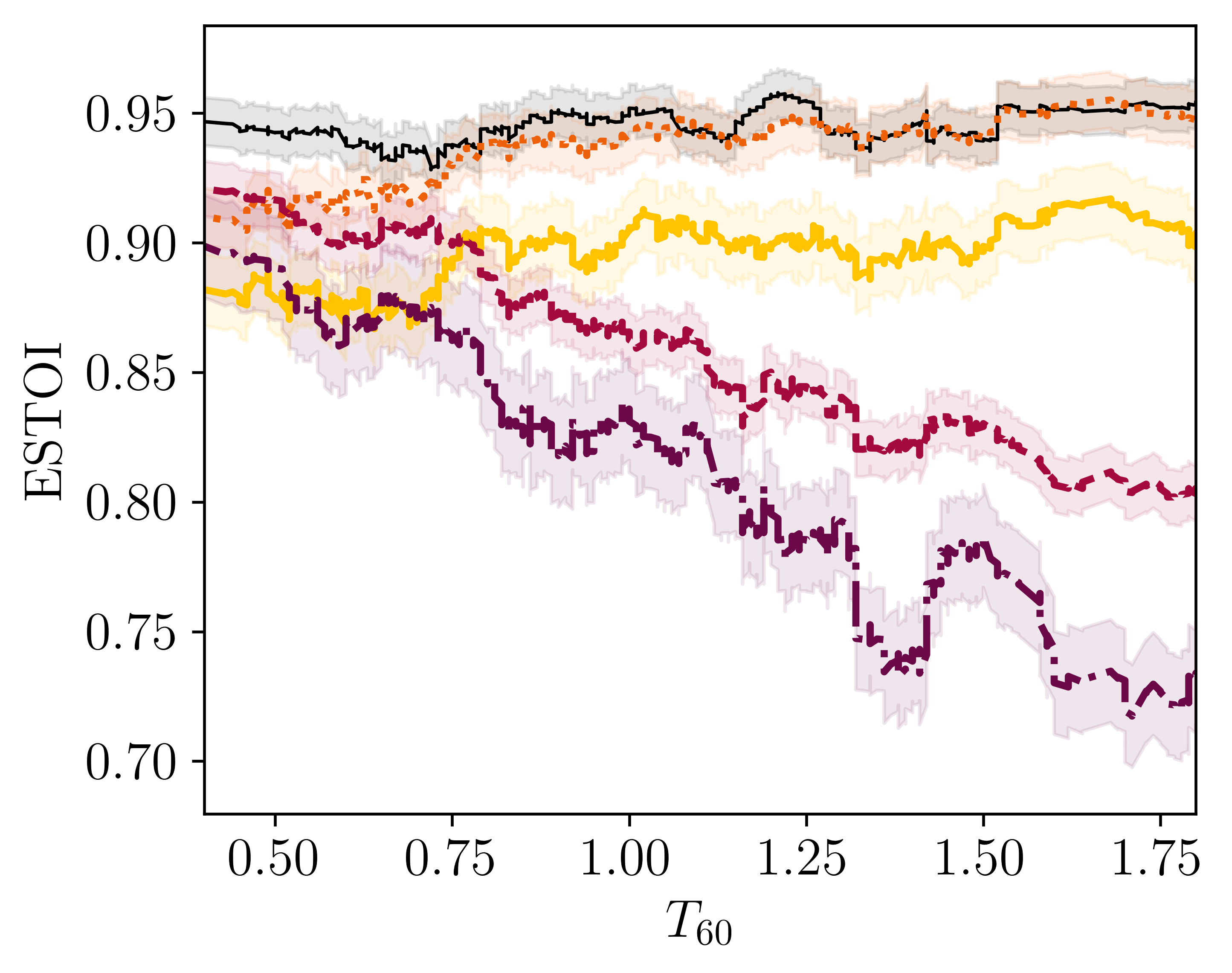}
    \caption{\protect\centering \textit{Speech metrics as a function of the $T_\mathrm{60}$ reverberation time. Colored areas cover half a standard deviation range.}
\vspace{-0.5em}}
    \label{fig:graph_t60}
\end{figure*}

%% file: plots/lineplot_measerr.tex
\newcommand{\w}{0.4\textwidth}
\newcommand{\h}{0.3\textwidth}
\newcommand{\xs}{0.05\textwidth}
\newcommand{\ys}{0.02\textwidth}
\newcommand{\lxs}{0.35\textwidth}

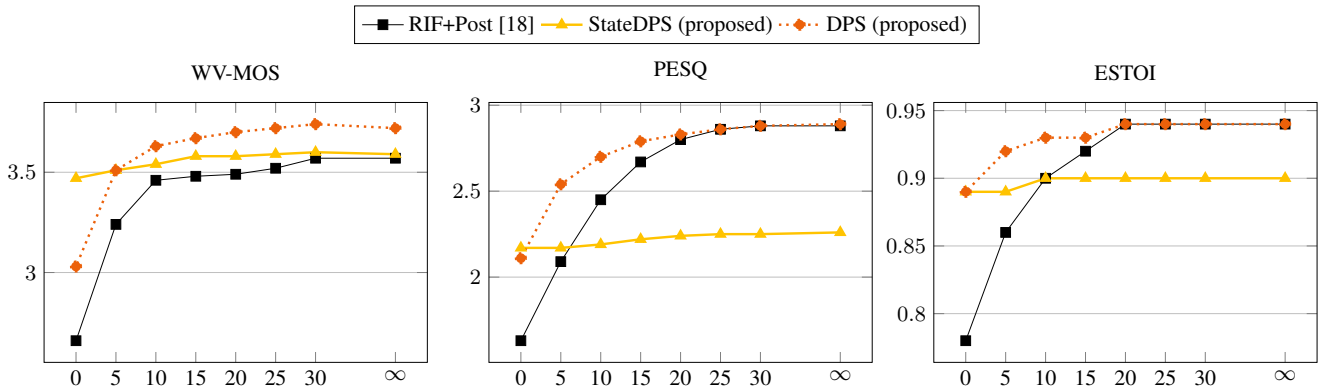
\begin{figure*}[t]
    \centering
\scalebox{0.92}{
\begin{tikzpicture}

\begin{axis}[title={WV-MOS}, name=wvmos,
ymajorgrids,
width=\w, height=\h,
xtick={0,5,10,15,20,25,30,40},
xticklabels={0,5,10,15,20,25,30,$\infty$},
legend style={
    at={(xticklabel cs:.5)},
    anchor=north,
    xshift=\lxs,
    yshift=5.6cm,
},
legend columns=5, mark=none,
]

\addplot+[kodrasi,, line width=0.25pt, mark=square*, mark options={fill=kodrasi}] table [x=SNR, y=WVMOS, col sep=comma] {kodrasi.csv};
\addplot+[statedps, line width=1pt, mark=triangle*, mark options={fill=statedps},] table [x=SNR, y=WVMOS, col sep=comma] {statedps.csv};
\addplot+[dps, line width=1pt, mark=*,dotted, mark options={fill=dps}] table [x=SNR, y=WVMOS, col sep=comma] {dps.csv};

\legend{RIF+Post \cite{Kodrasi2014}, StateDPS (proposed), DPS (proposed)}
\end{axis}

\begin{axis}[title={PESQ}, name=pesq, at={(wvmos.south east)},
ymajorgrids,
width=\w, height=\h, xshift=\xs,
xtick={0,5,10,15,20,25,30,40},
xticklabels={0,5,10,15,20,25,30,$\infty$},
]

\addplot+[kodrasi,, line width=0.25pt, mark=square*, mark options={fill=kodrasi}] table [x=SNR, y=PESQ, col sep=comma] {kodrasi.csv};
\addplot+[statedps, line width=1pt, mark=triangle*, mark options={fill=statedps},] table [x=SNR, y=PESQ, col sep=comma] {statedps.csv};
\addplot+[dps, line width=1pt, mark=*,dotted, mark options={fill=dps}] table [x=SNR, y=PESQ, col sep=comma] {dps.csv};

\end{axis}

\begin{axis}[title={ESTOI}, name=estoi, at={(pesq.south east)},
ymajorgrids,
width=\w, height=\h, xshift=\xs,
xtick={0,5,10,15,20,25,30,40},
xticklabels={0,5,10,15,20,25,30,$\infty$},
]

\addplot+[kodrasi,, line width=0.25pt, mark=square*, mark options={fill=kodrasi}] table [x=SNR, y=ESTOI, col sep=comma] {kodrasi.csv};
\addplot+[statedps, line width=1pt, mark=triangle*, mark options={fill=statedps},] table [x=SNR, y=ESTOI, col sep=comma] {statedps.csv};
\addplot+[dps, line width=1pt, mark=*,dotted, mark options={fill=dps}] table [x=SNR, y=ESTOI, col sep=comma] {dps.csv};

\end{axis}

\end{tikzpicture}
}
\vspace{-0.75em}
\caption{\centering \textit{Dereverberation performance under zero-mean Gaussian measurement noise.
Input SNR indicated on the horizontal axis in $\mathrm{dB}$.
}}
\label{fig:meas_err}
\vspace{-1.25em}
\end{figure*}

%% file: sections/exp.tex
\vspace{-0.5em}
\subsection{Data}
We generate the WSJ0+Reverb dataset as in \cite{Richter2022SGMSE++
} in a fashion resembling the WHAMR! dataset recipe \cite{Maciejewski2020whamr} by using clean speech data from the WSJ0 dataset and convolving each utterance with a simulated \ac{rir}. We use the \texttt{\small pyroomacoustics}
package \cite{Scheibler2018PyRoom} to simulate \acp{rir}. For each utterance, a reverberant room is modeled by sampling uniformly a $T_\mathrm{60}$ between 0.4 and 1.0 seconds and room dimensions in [5,15]$\times$[5,15]$\times$[2,6]~m. 
This results in an average \ac{drr} of -9$\,$dB and average \emph{measured} $T_\mathrm{60}$ of 0.91$\,$s. 
An anechoic (but auralized) version of the room is used to generate the reference clean speech, created using the same geometric parameters as the reverberant room but with the absorption coefficient set to 0.99.
\vspace{-1em}

\subsection{Hyperparameters and training configuration}
\label{sec:exp:hyperparameters}

\subsubsection{Data representation}
When training the unconditional score model, we use only the anechoic part of the generated WSJ0+Reverb data, as the model is supposed to learn the score over clean speech. Utterances are transformed using a \ac{stft} with a window size of 510 points, a hop length of 128 points and a square-root Hann window, at a sampling rate of $16$kHz. In contrast to \cite{Welker2022SGMSE,Richter2022SGMSE++}, no compression of the magnitude is used, in order to avoid instabilities when backpropagating small measurement errors through a non-linearity not differentiable around 0.
For training, segments of 256 \ac{stft} frames ($\approx$2s) are randomly extracted from the utterances and normalized by the maximum absolute value of the segment before feeding them to the network. 
Using publicly available code, the blind diffusion models SGMSE+ \cite{Richter2022SGMSE++} and StoRM \cite{Lemercier2022storm} are trained on the reverberant and anechoic speech datasets, as these methods are trained in a supervised setting. In comparison, the proposed method does not need any reverberant speech during training.
\vspace{-1em}

\subsubsection{Forward and reverse diffusion} 
We set the extremal noise levels for the diffusion schedule $g$ in \eqref{eq:forward-sde} to $\sigma_\mathrm{min}=$~0.05 and $\sigma_\mathrm{max}=$~0.5, and the terminal diffusion time to $T=1$. 
$50$ time steps are used for reverse diffusion with Algorithm~\ref{alg:inference}, which is adapted from the predictor-corrector scheme \cite{song2021sde} with probability flow ODE sampling and one step of annealed Langevin dynamics correction with step size of $r=$~0.4.

Tuning $\zeta^{'}(\tau)$ is quite difficult, as setting a high $\zeta^{'}$ leads to pre-echoes, feedback tones and other non-causality artifacts generated by the log-likelihood gradient. Using a low $\zeta^{'}$, however, puts too much emphasis on unconditional generation, therefore increasing the difference between the estimate and the original clean speech.
We notice that using the Annealed Langevin Dynamics corrector proposed in \cite{song2021sde} helps reduce the aforementioned artifacts and thus provides a more flexible tuning of $\zeta^{'}$.
We propose a saw-tooth schedule for $\zeta^{'}(\tau)$ where unconditional speech generation is promoted in the beginning of the reverse process and measurement importance is low towards the end of the process to avoid instabilities: 
\vspace{-0.5em}
\begin{equation}
    \zeta^{'}(\tau) = 
    \begin{cases}
    & \frac{2500 \tau}{0.9} \text{ if } \tau \leq 0.9 \\
    & \frac{2500 (1 - \tau)}{0.1} \text{ if } \tau \geq 0.9 \\ 
    \end{cases}
\end{equation}
\vspace{-2em}

\subsubsection{Network architecture} \label{sec:network}
The unconditional score network architecture is NCSN++M\cite{Lemercier2022analysing,Lemercier2022storm}, a lighter variant of the NCSN++ \cite{song2021sde} which uses $\sim$~27.8M parameters instead of the original 65M.
At each step $\tau$, the current state $\state$ real and imaginary channels are stacked and fed to the network, and the noise level $\sigma(\tau)$ is provided as a conditioner.
\vspace{-0.5em}

\subsubsection{Training configuration}
For training the unconditional score model, we use the Adam optimizer 
with a learning rate of $10^{-4}$ and an effective batch size of 16 for 300 epochs. 
We track an exponential moving average of the DNN weights with a decay of 0.999 to be used for sampling
as in \cite{Richter2022SGMSE++}.
A minimal diffusion time is set to $\tau_\epsilon=0.03$ during training to avoid singularities very close to $\tau = 0$.
\vspace{-1em}

\subsubsection{Evaluation metrics}
\label{sec:exp:eval}

For instrumental evaluation of the speech dereverberation performance, we use the intrusive \ac{pesq} \cite{Rix2001PESQ} and \ac{estoi} \cite{Jensen2016ESTOI} for assessment of speech quality and intelligibility respectively. We also use the non-intrusive WV-MOS \cite{Andreev2022Hifi++}\footnote{https://github.com/AndreevP/wvmos},
a DNN-based \ac{mos} approximation used in \cite{Lemercier2022analysing, Lemercier2022storm, Andreev2022Hifi++} for reference-free assessment of bandwidth extension and speech enhancement performance.

%% file: sections/results.tex
\subsection{Comparison to baselines}

In \figurename~\ref{fig:graph_t60}, we compare the proposed informed diffusion-based sampling schemes to the informed regularized inverse filtering plus post-processing baseline \cite{Kodrasi2014}, denoted in the following as RIF+Post. We further add comparisons to the blind dereverberation diffusion methods \cite{Richter2022SGMSE++, Lemercier2022storm}. Instrumental results are shown as a function of the input $T_\mathrm{60}$ reverberation time. While the performance of the blind dereverberation methods decreases as $T_\mathrm{60}$ increases, making the task more difficult, we observe that the informed methods exhibit consistent performance for all considered reverberation times. We notice that the proposed DPS method achieves better or comparable instrumental performance compared to the RIF+Post method \cite{Kodrasi2014}, while StateDPS performs overall poorer. This shows that using a denoised estimate to match the measurement model, as in the posterior mean approximation \eqref{eq:dps-likelihood},  increases the dereverberation performance as compared to the state approximation \eqref{eq:dolph-likelihood}.
Furthermore, the proposed DPS performed better in terms of subjective quality in informal listening tests: we refer the reader to the audio examples provided in our demo website (see link in abstract). 
As most diffusion schemes, the proposed (State)DPS methods and the baselines SGMSE+M and StoRM require multiple calls to the score network. Therefore, their computational burden is substantially superior to that of RIF+Post, which is a simple inverse filtering method with real-time capable post-processing.
\vspace{-0.5em}

\santiv
\subsection{Robustness to measurement error}

In \figurename~\ref{fig:meas_err}, we investigate the robustness of the informed dereverberation approaches to Gaussian measurement noise, added on top of the reverberant speech $\mathbf y$. 
We notice that the proposed DPS is significantly more robust to the introduced noise than the RIF+Post method. This is likely because the prior learned over clean speech by the score model helps gain robustness to mismatches in the measurement model. 
Informal experiments also show that the degradation of RIF+Post performance to real recorded environmental noise is dramatic, while DPS maintains a very high dereverberation performance and simply lets noise pass through. This shows that the proposed method  is much more reliable in realistic scenarios where various noise sources arise, as the remaining noise after DPS dereverberation can easily be removed by a post-processing stage.

\vspace{-0.5em}
\santiv
\subsection{Extension to blind dereverberation}

An important 
aspect of the presented work is that the proposed diffusion posterior sampling technique for informed dereverberation can be extended to blind dereverberation using \cite{chung_parallel_2022}. In \cite{chung_parallel_2022} a framework for joint estimation of the blurring kernel and target image is developed using parallel diffusion processes. Our work lays the ground for future adaptation of \cite{chung_parallel_2022} to jointly estimate the \ac{rir} and clean speech in subsequent work. The method we have presented here is interpretable due to the explicit forward model, in contrast to blind dereverberation approaches such as \cite{Richter2022SGMSE++,Lemercier2022storm}. If successfully implemented, the future method would combine this interpretability with a generative estimation of \acp{rir}. This would furthermore dispose of the need for plug-in \ac{rir} estimators in blind scenarios, which the baseline method \cite{Kodrasi2014} in contrast requires.

%% file: sections/conclusion.tex
We have presented a single-channel informed dereverberation method based on diffusion models. The proposed method uses a clean speech prior parameterized by a score model as well as a log-likelihood approximation to generate anechoic speech that fits the measurement model.
The approach outperforms an existing state-of-the-art frequency-domain method in terms of robustness to both white Gaussian and real environmental measurement noises. 
One of the introduced sampling schemes also largely outperforms existing diffusion-based blind dereverberation methods for long reverberation times. 
The work at hand lays ground to an interpretable extension to blind dereverberation using joint estimation of the \ac{rir} and anechoic speech with diffusion models.

%% file: ms.bbl
\begin{thebibliography}{10}
\providecommand{\url}[1]{#1}
\def\UrlFont{\rmfamily}
\providecommand{\newblock}{\relax}
\providecommand{\bibinfo}[2]{#2}
\providecommand\BIBentrySTDinterwordspacing{\spaceskip=0pt\relax}
\providecommand\BIBentryALTinterwordstretchfactor{4}
\providecommand\BIBentryALTinterwordspacing{\spaceskip=\fontdimen2\font plus
\BIBentryALTinterwordstretchfactor\fontdimen3\font minus
  \fontdimen4\font\relax}
\providecommand\BIBforeignlanguage[2]{{%
\expandafter\ifx\csname l@#1\endcsname\relax
\typeout{** WARNING: IEEEtran.bst: No hyphenation pattern has been}%
\typeout{** loaded for the language `#1'. Using the pattern for}%
\typeout{** the default language instead.}%
\else
\language=\csname l@#1\endcsname
\fi
#2}}

\bibitem{Naylor2011}
P.~A. Naylor and N.~D. Gaubitch, \emph{Speech Dereverberation}.\hskip 1em plus
  0.5em minus 0.4em\relax Springer, 2011, vol.~59.

\bibitem{gerkmann2018book_chapter}
T.~Gerkmann and E.~Vincent, \emph{Spectral Masking and Filtering}.\hskip 1em
  plus 0.5em minus 0.4em\relax John Wiley \& Sons, 2018.

\bibitem{wang2018supervised}
D.~Wang and J.~Chen, ``Supervised speech separation based on deep learning: An
  overview,'' \emph{IEEE Trans. Audio, Speech, Language Proc.}, vol.~26,
  no.~10, pp. 1702--1726, 2018.

\bibitem{Williamson2017j}
D.~S. Williamson and D.~Wang, ``Time-frequency masking in the complex domain
  for speech dereverberation and denoising,'' \emph{IEEE/ACM Trans. Audio,
  Speech, Language Proc.}, vol.~25, no.~7, pp. 1492--1501, 2017.

\bibitem{Ernst2019}
O.~Ernst, S.~E. Chazan, S.~Gannot, and J.~Goldberger, ``Speech dereverberation
  using fully convolutional networks,'' in \emph{Proc. Euro. Signal Proc. Conf.
  (EUSIPCO)}, Sept. 2019.

\bibitem{Han2017}
K.~Han, Y.~Wang, D.~Wang, W.~S. Woods, I.~Merks, and T.~Zhang, ``Learning
  spectral mapping for speech dereverberation and denoising,'' \emph{IEEE/ACM
  Trans. Audio, Speech, Language Proc.}, vol.~23, no.~6, pp. 982--992, 2015.

\bibitem{ho2020denoising}
J.~Ho, A.~Jain, and P.~Abbeel, ``Denoising diffusion probabilistic models,'' in
  \emph{Neural Information Proc. Systems (NIPS)}, Dec. 2020.

\bibitem{song2021sde}
Y.~Song, J.~Sohl-Dickstein, D.~P. Kingma, A.~Kumar, S.~Ermon, and B.~Poole,
  ``Score-based generative modeling through stochastic differential
  equations,'' in \emph{Int. Conf. Learning Repr. (ICLR)}, May 2021.

\bibitem{Richter2022SGMSE++}
J.~Richter, S.~Welker, J.-M. Lemercier, B.~Lay, and T.~Gerkmann, ``Speech
  enhancement and dereverberation with diffusion-based generative models,''
  \emph{IEEE Trans. Audio, Speech, Language Proc.}, pp. 1--13, 2023.

\bibitem{Lemercier2022storm}
J.-M. Lemercier, J.~Richter, S.~Welker, and T.~Gerkmann, ``{S}to{RM}: A
  diffusion-based stochastic regeneration model for speech enhancement and
  dereverberation,'' \emph{arXiv}, Dec. 2022.

\bibitem{lu2022conditional}
Y.-J. Lu, Z.-Q. Wang, S.~Watanabe, A.~Richard, C.~Yu, and Y.~Tsao,
  ``Conditional diffusion probabilistic model for speech enhancement,'' in
  \emph{IEEE Int. Conf. Acoustics, Speech, Signal Proc. (ICASSP)}, June 2022.

\bibitem{Neely1979Invertibility}
S.~T. Neely and J.~B. Allen, ``Invertibility of a room impulse response,''
  \emph{The Journal of the Acoustical Society of America}, vol.~66, no.~1, pp.
  165--169, 07 1979.

\bibitem{Miyoshi1988MINT}
M.~Miyoshi and Y.~Kaneda, ``Inverse filtering of room acoustics,'' \emph{IEEE
  Trans. Audio, Speech, Language Proc.}, vol.~36, no.~2, pp. 145--152, 1988.

\bibitem{Hikichi2007Inverse}
T.~Hikichi, M.~Delcroix, and M.~Miyoshi, ``Inverse filtering for speech
  dereverberation less sensitive to noise and room transfer function
  fluctuations,'' \emph{EURASIP J. Adv. Sig. Proc.}, vol. 2007, Dec. 2007.

\bibitem{Mourjopoulos1982Comparative}
J.~Mourjopoulos, P.~Clarkson, and J.~Hammond, ``A comparative study of
  least-squares and homomorphic techniques for the inversion of mixed phase
  signals,'' in \emph{IEEE Int. Conf. Acoustics, Speech, Signal Proc.
  (ICASSP)}, June 1982.

\bibitem{Mertins2010Shortening}
A.~Mertins, T.~Mei, and M.~Kallinger, ``Room impulse response
  shortening/reshaping with infinity- and $p$-norm optimization,'' \emph{IEEE
  Trans. Audio, Speech, Language Proc.}, vol.~18, no.~2, pp. 249--259, 2010.

\bibitem{Schepker2022Robust}
H.~Schepker, F.~Denk, B.~Kollmeier, and S.~Doclo, ``Robust single- and
  multi-loudspeaker least-squares-based equalization for hearing devices,''
  \emph{EURASIP J. Aud. Speech and Mus. Proc.}, vol. 2022, pp. 1--14, 06 2022.

\bibitem{Kodrasi2014}
I.~Kodrasi, T.~Gerkmann, and S.~Doclo, ``Frequency-domain single-channel
  inverse filtering for speech dereverberation: Theory and practice,'' in
  \emph{IEEE Int. Conf. Acoustics, Speech, Signal Proc. (ICASSP)}, May 2014.

\bibitem{Welker2022SGMSE}
S.~Welker, J.~Richter, and T.~Gerkmann, ``Speech enhancement with score-based
  generative models in the complex {STFT} domain,'' in \emph{Interspeech},
  Sept. 2022.

\bibitem{Lemercier2022analysing}
J.-M. Lemercier, J.~Richter, S.~Welker, and T.~Gerkmann, ``Analysing
  discriminative versus diffusion generative models for speech restoration
  tasks,'' in \emph{IEEE Int. Conf. Acoustics, Speech, Signal Proc. (ICASSP)},
  June 2023.

\bibitem{anderson1982reverse}
B.~D. Anderson, ``Reverse-time diffusion equation models,'' \emph{Stochastic
  Processes and their Applications}, vol.~12, no.~3, pp. 313--326, 1982.

\bibitem{vincent2011connection}
P.~Vincent, ``A connection between score matching and denoising autoencoders,''
  \emph{Neural Computation}, vol.~23, no.~7, pp. 1661--1674, 2011.

\bibitem{chung_diffusion_2022}
H.~Chung, J.~Kim, M.~T. Mccann, M.~L. Klasky, and J.~C. Ye, ``Diffusion
  posterior sampling for general noisy inverse problems,'' \emph{Int. Conf.
  Learning Repr. (ICLR)}, May 2023.

\bibitem{moliner_solving_2022}
E.~Moliner, J.~Lehtinen, and V.~Välimäki, ``Solving audio inverse problems
  with a diffusion model,'' in \emph{IEEE Int. Conf. Acoustics, Speech, Signal
  Proc. (ICASSP)}, June 2023.

\bibitem{Efron2011}
B.~Efron, ``Tweedie’s formula and selection bias,'' \emph{Journal of the
  American Statistical Association}, vol. 106, no. 496, pp. 1602--1614, 2011.

\bibitem{shoushtari2022dolph}
S.~Shoushtari, J.~Liu, and U.~S. Kamilov, ``{DOLPH}: Diffusion models for phase
  retrieval,'' \emph{arXiv}, Nov. 2022.

\bibitem{Maciejewski2020whamr}
M.~Maciejewski, G.~Wichern, E.~McQuinn, and J.~L. Roux, ``{WHAMR!}: {N}oisy and
  reverberant single-channel speech separation,'' in \emph{IEEE Int. Conf.
  Acoustics, Speech, Signal Proc. (ICASSP)}, 2020.

\bibitem{Scheibler2018PyRoom}
R.~Scheibler, E.~Bezzam, and I.~Dokmanic, ``Pyroomacoustics: A python package
  for audio room simulation and array processing algorithms,'' in \emph{IEEE
  Int. Conf. Acoustics, Speech, Signal Proc. (ICASSP)}, Apr. 2018.

\bibitem{Rix2001PESQ}
A.~Rix, J.~Beerends, M.~Hollier, and A.~Hekstra, ``Perceptual evaluation of
  speech quality ({PESQ})-a new method for speech quality assessment of
  telephone networks and codecs,'' in \emph{IEEE Int. Conf. Acoustics, Speech,
  Signal Proc. (ICASSP)}, May 2001.

\bibitem{Jensen2016ESTOI}
J.~Jensen and C.~Taal, ``An algorithm for predicting the intelligibility of
  speech masked by modulated noise maskers,'' \emph{IEEE/ACM Trans. Audio,
  Speech, Language Proc.}, vol.~24, no.~11, pp. 2009--2022, 2016.

\bibitem{Andreev2022Hifi++}
P.~Andreev, A.~Alanov, O.~Ivanov, and D.~Vetrov, ``Hifi++: a unified framework
  for bandwidth extension and speech enhancement,'' in \emph{IEEE Int. Conf.
  Acoustics, Speech, Signal Proc. (ICASSP)}, June 2023.

\bibitem{chung_parallel_2022}
H.~Chung, J.~Kim, S.~Kim, and J.~C. Ye, ``Parallel diffusion models of operator
  and image for blind inverse problems,'' \emph{IEEE/CVF Conf. on Computer
  Vision and Pattern Recognition (CVPR)}, June 2023.

\end{thebibliography}
